\documentclass[aps,prl,reprint,superscriptaddress,showpacs,preprintnumbers,amsmath,amssymb]{revtex4-2}
\usepackage{graphicx}

\usepackage{hyperref}

\usepackage[all]{hypcap}

\usepackage{dcolumn}   % needed for some tables
\usepackage{bm}        % for math
\usepackage{amssymb, amsmath, amsfonts}   % for math
\usepackage{xcolor}
\usepackage{ulem}
\DeclareGraphicsExtensions{{.pdf}}
\begin{document}
\newcommand{\noter}[1]{{\color{red}{#1}}}
\newcommand{\noteb}[1]{{\color{blue}{#1}}}
\newcommand{\field}{\left( \boldsymbol{r}\right)}
\newcommand{\paren}[1]{\left({#1}\right)}
\newcommand{\vect}[1]{\boldsymbol{#1}}
\newcommand{\uvect}[1]{\tilde{\boldsymbol{#1}}}
\newcommand{\vdot}[1]{\dot{\boldsymbol{#1}}}
\newcommand{\vder}{\boldsymbol{\nabla}}
\newcommand{\be}{\begin{equation}}
\newcommand{\ee}{\end{equation}}
\newcommand{\bea}{\begin{equnaray}}
\newcommand{\eea}{\end{equnaray}}
\newcommand{\ba}{\begin{align}}
\newcommand{\ea}{\end{align}}
\newcommand{\ave}[1]{\left\langle {#1} \right\rangle}
\widetext

\title{Dense HeLa cell monolayers remain liquid-like despite strong crowding}

\author{Suravi Pal}
\affiliation{Life Science Center for Survival Dynamics, University of Tsukuba, Ibaraki 305-8577, Japan}
\affiliation{D3 Center, The University of Osaka, Toyonaka, Osaka 560-0043, Japan}

\author{Nen Saito}
\affiliation{Life Science Center for Survival Dynamics, University of Tsukuba, Ibaraki 305-8577, Japan}

\author{Takeshi Kawasaki}
\thanks{These authors jointly supervised this work.}
\affiliation{D3 Center, The University of Osaka, Toyonaka, Osaka 560-0043, Japan}
\affiliation{Department of Physics, The University of Osaka, Toyonaka, Osaka 560-0043, Japan}
\email{kawasaki.takeshi.d3c@osaka-u.ac.jp}

\author{Hiroyuki Ebata}
\thanks{These authors jointly supervised this work.}
\affiliation{Department of Earth and Space Science, The University of Osaka, Toyonaka, Osaka 560-0043, Japan}
\email{ebata@ess.sci.osaka-u.ac.jp}

\begin{abstract}
Collective dynamics in dense cell monolayers are governed by the interplay between crowding and cellular motility. Although increasing density can slow cellular motion and promote glass-like behaviour, the dynamical state of dense HeLa monolayers remains unclear. Here, we combine in vitro time-lapse imaging of HeLa cell monolayers with simulations of a deformable active-cell model to examine how cell density and motility regulate collective relaxation. Within the experimentally accessible density and time ranges, untreated HeLa monolayers remain liquid-like: structural relaxation progressively slows down with increasing density but remains observable throughout the investigated range. Under low-nutrient conditions, cell motility is strongly reduced, and structural relaxation becomes substantially slower. To elucidate the mechanisms underlying these experimental observations, we further performed simulations using a deformable-cell model. The model qualitatively reproduces the density-dependent increase in structural relaxation time and further shows that reducing self-propulsion promotes long-lived caging dynamics at high packing fractions. These results show that dense HeLa monolayers can sustain slow, heterogeneous, yet relaxing collective dynamics under untreated conditions, and indicate that persistent cellular motility is an important factor in maintaining structural relaxation at high density, which may provide insight into the metastatic potential of cancer cells.
\end{abstract}

\maketitle
\section{Introduction}
% Cancer
Cancer is a malignant disease in which cells accumulate genetic mutations that confer unlimited proliferative and metastatic capacity, and elucidating the mechanisms underlying metastasis remains one of the central challenges in cancer therapy~\cite{Wirtz2011NatRevCancer}.
Recent studies have increasingly shown that collective cell migration and the mechanical properties of tumour tissue play a critical role in the metastatic process, highlighting the need for a tissue-scale physical understanding that goes beyond the molecular and genetic characterization of individual cells~\cite{Friedl2009NatRevMolCellBiol,Oswald2017J.Phys.D:Appl.Phys.}.

Cell assemblies can be described within the same physical framework as colloidal suspensions and granular matter, since individual cells occupy finite volumes, possess well-defined shapes, and move under mutual crowding~\cite{Angelini2011PNAS}.
This framework provides a natural starting point for understanding how density and mechanical interactions regulate collective cell motion at the tissue scale~\cite{Trepat2018NaturePhys}. 
With increasing density, structural relaxation can slow dramatically with only modest structural changes and become dynamically heterogeneous, with an approach to a glass transition~\cite{Kob1995Phys.Rev.E,Yamamoto1998Phys.Rev.E,Debenedetti2001Nature,Berthier2011Rev.Mod.Phys.,Cavagna2009Phys.Rep.,Janssen2018Front.Phys.}.

Indeed, densely packed cancer cells in tumour explants exhibit dynamically heterogeneous motion, with relatively mobile and immobile regions coexisting within the same tissue, suggesting that concepts from glass physics may provide a useful framework for understanding their collective dynamics.~\cite{Gottheil2023Phys.Rev.X,Staneva2019JCellSci}.

% active matter and active glasses
However, the glass-transition framework was originally developed for passive particle systems driven by thermal fluctuations and requires extension when applied to self-propelled particle systems---active matter---such as living cells, which continuously consume energy to generate motion and mechanical forces~\cite{Marchetti2013Rev.Mod.Phys.,Bechinger2016Rev.Mod.Phys.}. In active particle systems, theoretical and numerical studies have established that, in addition to density and temperature, the magnitude of self-propulsion and its persistence time act as independent control parameters governing structural relaxation~\cite{Henkes2011Phys.Rev.E,Fily2012Phys.Rev.Lett.}. Specifically, increasing self-propulsion enables particle rearrangements even at densities where a passive system would be dynamically arrested, markedly reducing the structural relaxation time and shifting the glass-transition line within the density--activity plane~\cite{Ni2013Nat.Commun.,Berthier2014Phys.Rev.Lett.,Janssen2019J.Phys.:Condens.Matter,Sadhukhan2024Eur.Phys.J.Spec.Top.}. In other words, in active matter, the boundary between glassy arrest and fluid-like flow is set not by density alone but jointly by density and activity~\cite{Berthier2014Phys.Rev.Lett.,Bi2016Phys.Rev.X}. 

% shape induced glasses/jamming
The concepts of glassy dynamics and active fluidization have also been applied to dense biological tissues. Experiments on confluent epithelial monolayers have shown pronounced dynamical slowing and collective jamming with increasing cell density~\cite{Garcia2015Proc.Natl.Acad.Sci.,Nnetu2013SoftMatter}. Vertex-model studies, in particular, have shown that cell shape provides an additional degree of freedom governing tissue rigidity beyond density alone: a dimensionless shape index, which quantifies cell-shape anisotropy, acts as an important control parameter, such that decreasing the shape index below a critical value can drive the tissue from fluid-like to solid-like behaviour even in the idealized limit where density is held fixed~\cite{Bi2015NatPhys}. Motility further modifies this rigidity landscape and can drive glass and jamming transitions at fixed density~\cite{Bi2016Phys.Rev.X}. These results show that, in living tissues, cell shape provides an additional control variable beyond density and activity~\cite{Atia2018NaturePhys}.

% density/packing fraction + shape
In the vertex-model framework, however, the preferred cell shape is typically encoded through an externally specified target-shape parameter rather than emerging directly from the underlying cell-level interactions~\cite{Bi2015NatPhys}. Moreover, conventional vertex models generally describe confluent tissues and therefore do not naturally accommodate variations in packing fraction away from confluence. This limitation may be particularly relevant to cancer cell collectives with weak cell--cell adhesion, where local cell density can decrease near the tumour boundary from which invasion occurs~\cite{Kang2021iScience,Net2024iScience}. Models that instead represent individual cells as deformable, non-polygonal objects allow cell shape to evolve dynamically in response to cell--cell interactions and crowding~\cite{Lober2015SciRep,Saito2024Sci.Adv.}. In particular, the Fourier-contour-cell model (mentioined as the deformable-cell model later onwards) allows packing fraction, cell deformability, and activity to be varied independently, making it well suited for disentangling their respective effects on collective dynamics~\cite{Saito2024Sci.Adv.}. These features make the deformable-cell model a promising framework for describing collective cancer cell migration, although its applicability to such systems is yet to be examined.

% To the cancer study
% Aim
Despite these advances, it remains unclear how quantitatively the activity-controlled glass scenario applies to actual cancer-cell monolayers. In particular, direct comparisons between experiment and theory that independently assess the effects of crowding and cellular activity on structural relaxation remain limited. This question is unresolved even for HeLa cells, one of the most extensively studied human cancer cell lines~\cite{Masters2002NatRevCancer}: it is not known whether increasing density alone can drive dense HeLa monolayers toward glass-like arrest, or whether a reduction in cellular activity is required to produce pronounced slower collective dynamics. 
To address this question, we combine \textit{in vitro} time-lapse imaging of dense HeLa monolayers with simulations of the deformable-cell model~\cite{Saito2024Sci.Adv.} to determine how crowding and cellular activity jointly control cell morphology, dynamical heterogeneity, and structural relaxation---three observables that together characterize whether a cell layer remains fluid-like or approaches dynamical arrest.

% Contents of this study
In this study, we first identify a representative simulation condition by comparing cell shape and single-cell displacement statistics between experiment and simulation. Afterward, we quantify how increasing density affects collective relaxation and test whether crowding alone is sufficient to produce dynamical arrest. Finally, by reducing cellular activity experimentally and varying self-propulsion independently in the model, we examine whether this can drive dense HeLa monolayers toward a glass-like state. This combined experiment--simulation approach enables us to distinguish crowding-induced slowing from activity-controlled arrest and to establish a quantitative correspondence between the collective dynamics of HeLa monolayers and a physically interpretable deformable-cell model. By clarifying the physical factors that govern collective arrest in a widely used cancer-cell line, this work provides a framework for future studies exploring possible links between tissue-scale fluidity, cancer-cell migration, and metastatic potential.

%%%
\section{Results}
We observed dense HeLa cell monolayers by time-lapse microscopy and extracted individual cell contours and trajectories using a deep-learning-assisted cell segmentation and tracking method to quantify cell morphology and collective relaxation. 

The experimental observations were compared with simulations of the polydisperse active deformable-cell model (Fourier-contour-cell model) developed by Saito and Ishihara~\cite{Saito2024Sci.Adv.}.
In this model, self-propelled cells deform through the interplay between steric interactions and membrane tension. We evaluated the same structural and dynamical observables in both experiments and simulations, allowing us to compare their collective behaviours within a common framework.

\subsection{Collective migration in dense HeLa cancer-cell monolayers}

\begin{figure*}[t]
\centering
\includegraphics[width=0.95\textwidth]{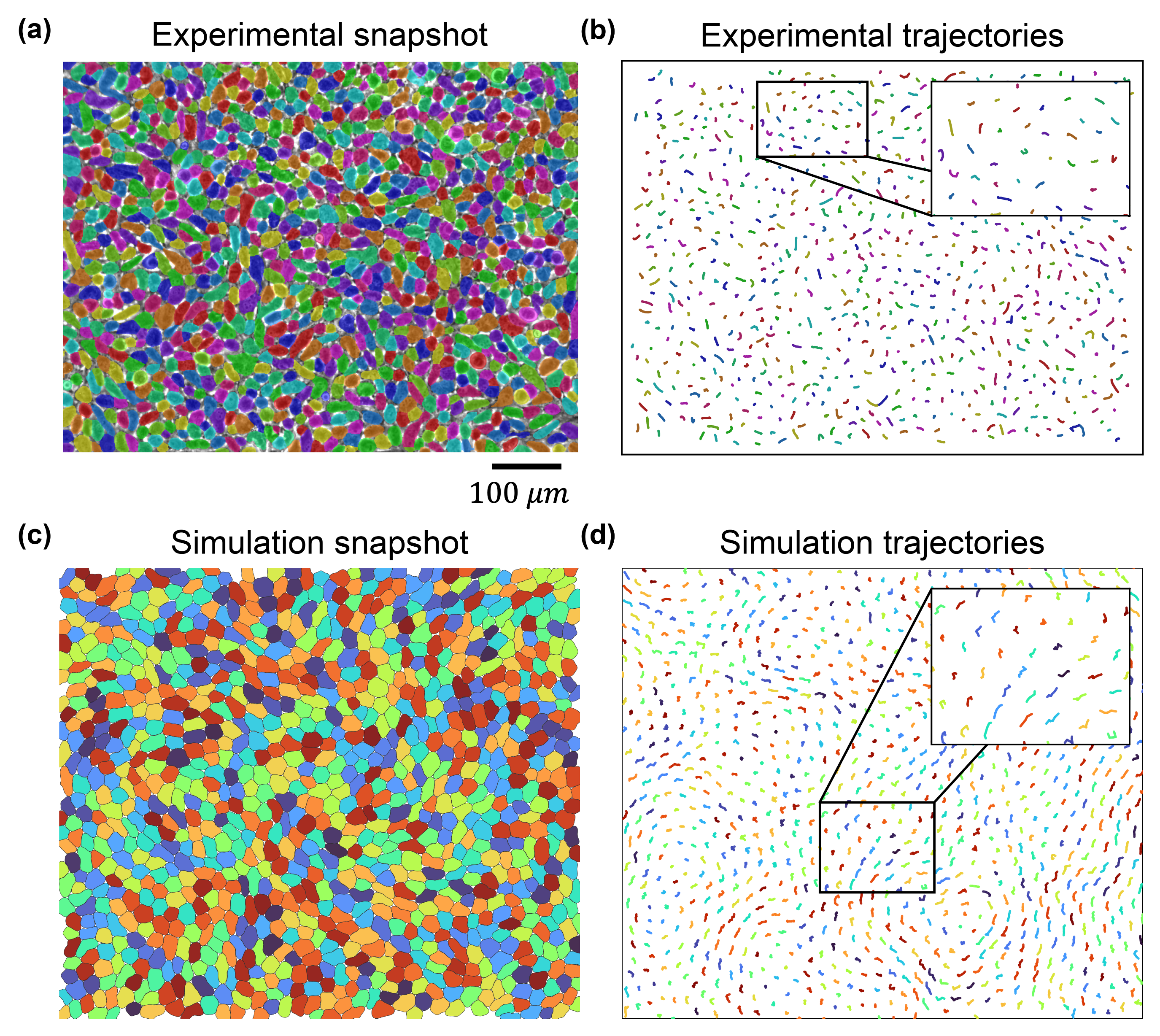}
\caption{
\textbf{Collective organization and migration in dense HeLa cancer-cell monolayers : experiment and simullation}.
\textbf{a,} Representative phase-contrast image of a confluent HeLa monolayer experiment, with individual cells identified by overlaid colour-coded segmentation masks. The cell density is approximately $2.5\times10^{3}$ cells/mm$^{2}$. Scale bar, $100~\mu$m.
\textbf{b,} Experimentally measured cell trajectories over approximately one structural relaxation time at the same cell density. The boxed region is enlarged in the inset to highlight local cell-scale displacements.
\textbf{c,} Representative steady-state snapshot from the deformable-cell model at packing fraction $\phi=0.97$, self-propulsion speed $v_0=0.03$, and deformability parameter $\eta=0.01$. Individual simulated cells are shown in random colours according to particle identity.
\textbf{d,} Simulated cell trajectories accumulated over one structural relaxation-time window for the same parameter set as in \textbf{c}. The trajectories are shown within the simulation box, and the boxed region is magnified in the inset to illustrate local particle-scale rearrangements. 
}
\label{fig:fig1}
\end{figure*}

Dense cancer-cell monolayers exhibit collective migration governed by the interplay among cellular motility, mechanical interactions, and geometric confinement. We first compared representative configurations and trajectories of confluent HeLa monolayers with those obtained from the deformable-cell model. Figure~\ref{fig:fig1}\textbf{a} shows a segmented phase-contrast image of a dense HeLa monolayer. The cells adopt heterogeneous, irregular shapes while collectively forming an approximately space-filling layer. 

To visualize the collective dynamics, we tracked individual cells over approximately one structural relaxation time. The resulting trajectories (Fig.~\ref{fig:fig1}\textbf{b}) show appreciable displacements within the crowded monolayer. Neighbouring cells tend to move in correlated directions, visually suggesting collective multicellular rearrangements. 

We next examined whether these qualitative features could be reproduced by the model, which incorporates cell deformation, self-propulsion, and steric interactions. A representative simulated configuration (Fig.~\ref{fig:fig1}\textbf{c}) exhibits irregular cell shapes and confluent packing similar to those observed experimentally. The corresponding trajectories (Fig.~\ref{fig:fig1}\textbf{d}) also display heterogeneous and spatially correlated multicellular motion. This qualitative correspondence motivates the quantitative comparison of cell morphology, displacement statistics, and structural relaxation as presented below.

\subsection{Cell shape and displacement statistics identify a representative simulation condition}

We next identified a representative simulation condition for comparison with the HeLa monolayer by examining how cell morphology depends on packing fraction, self-propulsion speed, and deformability. Cell shape was quantified using the dimensionless shape index
$p_i=P_i/\sqrt{A_i}$, where $P_i$ and $A_i$ denote perimeter and area of the cell $i$, respectively (see Methods).

\begin{figure*}[ht]
\centering
\includegraphics[width=0.79\textwidth]{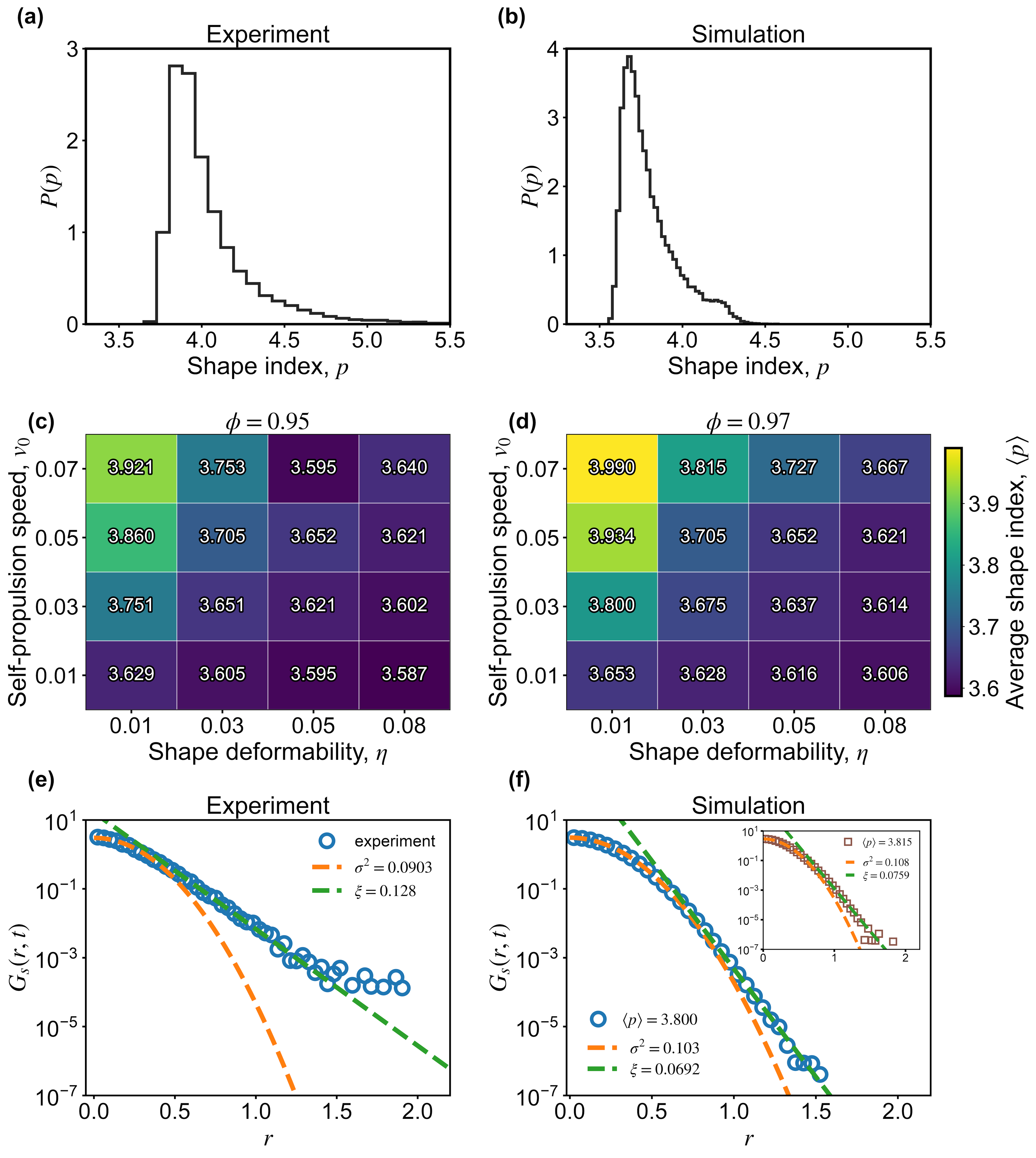}
\caption{
\textbf{Cell morphology and single-cell displacement statistics in experiments and simulations.}
\textbf{a,} Representative experimental probability-density distribution of the cell shape index, $P(p)$.
\textbf{b,} Simulated $P(p)$ at $\phi=0.97$, $v_0=0.03$, and $\eta=0.01$.
\textbf{c,d,} Average cell shape index, $\langle p\rangle$, obtained from the deformable-cell model as a function of self-propulsion speed $v_0$ and shape deformability $\eta$ at $\phi=0.95$ (\textbf{c}) and $\phi=0.97$ (\textbf{d}). The value displayed in each grid cell is the corresponding average shape index. Increasing $v_0$ and decreasing $\eta$ lead to stronger cell deformation and hence larger $\langle p\rangle$.
\textbf{e,} Representative radial self-part of the van Hove correlation function, $G_{\mathrm{s}}(r,t)$, measured experimentally at lag time $t=\tau_\alpha$, where $\tau_\alpha$ is the structural relaxation time. Open circles denote the experimental data. The distance $r$ is normalized by the mean cell diameter, $d$. The orange dashed line is a Gaussian fit to the central-displacement region, with variance $\sigma^{2}=0.0903$, and the green dashed line is an exponential fit to the large-displacement tail, with decay length $\xi=0.128$. 
\textbf{f,} Corresponding displacement distribution obtained from the simulation at $\phi=0.97$, $v_0=0.03$, and $\eta=0.01$, evaluated at $t=\tau_\alpha$. This state has $\langle p\rangle=3.800$, with a Gaussian-core variance $\sigma^{2}=0.103$ and an exponential-tail decay length $\xi=0.0692$. The upper-right inset shows a second morphology-matched state at the same packing fraction, with $v_0=0.07$, $\eta=0.03$, $\langle p\rangle=3.815$, $\sigma^{2}=0.108$, and $\xi=0.0759$. 
\textbf{a, e,} The experimental cell density was approximately $2.5\times10^{3}$ cells~mm$^{-2}$.
}
\label{fig:fig2}
\end{figure*}

Figure~\ref{fig:fig2}\textbf{a} shows the experimental shape-index distribution $P(p)$ at a cell density of approximately $2.5\times10^{3}$ cells/mm$^{2}$. The distribution is asymmetric, with a maximum near $p\simeq3.9$ and a broad tail towards larger values, reflecting the coexistence of relatively compact and more elongated cells within the monolayer. For comparison, at $\phi=0.97$, $v_0=0.03$, and $\eta=0.01$, the simulated distribution exhibits a similar single-peaked, right-skewed form (Fig.~\ref{fig:fig2}\textbf{b}). Although the detailed width and large-$p$ tail differ from those measured experimentally, the model captures key qualitative features of the observed cell-shape distribution.

%Next, we examined how the shape index depends on self-propulsion and deformability.
Figures~\ref{fig:fig2}\textbf{c} and \ref{fig:fig2}\textbf{d} show a phase diagram of average shape index $\langle p\rangle$ in the $(v_0,\eta)$ plane at $\phi=0.95$ and $0.97$, respectively. At both packing fractions, $\langle p\rangle$ generally increases with increasing $v_0$ and decreasing $\eta$. Because a smaller $\eta$ corresponds to a weaker energetic penalty for contour elongation, cells deform more readily in this regime. Increasing the packing fraction from $\phi=0.95$ to $\phi=0.97$ also moderately increases $\langle p\rangle$ for several parameter combinations, consistent with stronger geometric confinement and more frequent cell--cell interactions at higher density. 
The experimental average shape index is approximately $3.9$--$4.0$ at high density. Although parameter combinations with high $v_0$ and low $\eta$ yield average values closer to this range, their $P(p)$ become bimodal, qualitatively differing from the single-peaked experimental distribution (Fig.~S1). 
In contrast, $(\phi,v_0,\eta)=(0.97,0.03,0.01)$ and $(0.97,0.07,0.03)$ yielded single-peaked $P(p)$ qualitatively similar to the experimental distribution, despite somewhat lower average shape indices. We therefore evaluated these conditions further by comparing single-cell displacement statistics using the self-part of the van Hove correlation function.

Figure~\ref{fig:fig2}\textbf{e} shows the experimental displacement distribution $G_{\mathrm{s}}(r,t)$ at a lag time comparable to the structural relaxation time~(see Methods). The central region is approximately Gaussian, whereas the large-displacement region decays more slowly and is better described by an exponential tail. This deviation from a single Gaussian form indicates heterogeneous cell dynamics: most cells undergo relatively small displacements, whereas a smaller fraction undergoes substantially larger rearrangements over the same time interval. A similar combination of a Gaussian-like core and an exponential tail is observed in the simulations (Fig.~\ref{fig:fig2}\textbf{f}). 
Both $(\phi,v_0,\eta)=(0.97,0.03,0.01)$ and $(0.97,0.07,0.03)$ (inset) qualitatively reproduce the broad, non-Gaussian form of the experimental displacement distribution. Thus, different combinations of activity and deformability can reproduce key morphological and dynamical features of the experimental monolayer.

The complete $P(p)$ for the candidate parameter sets are compared with the experimental distribution in the Supplementary Information. 
Together, the shape-index and displacement distributions substantially constrain the range of suitable parameters but do not uniquely determine a single parameter set. For the subsequent analysis, we used $(\phi,v_0,\eta)=(0.97,0.03,0.01)$ as a representative reference condition.

\subsection{Crowding slows structural relaxation without observable dynamical arrest}

Using $v_0=0.03$ and $\eta=0.01$ as representative simulation parameters, we examined whether the model reproduces the evolution of collective relaxation with increasing crowding. We compared self-intermediate scattering functions and structural relaxation times over a range of experimental cell densities and simulated packing fractions. 

\begin{figure*}[ht]
\centering
\includegraphics[width=0.93\textwidth]{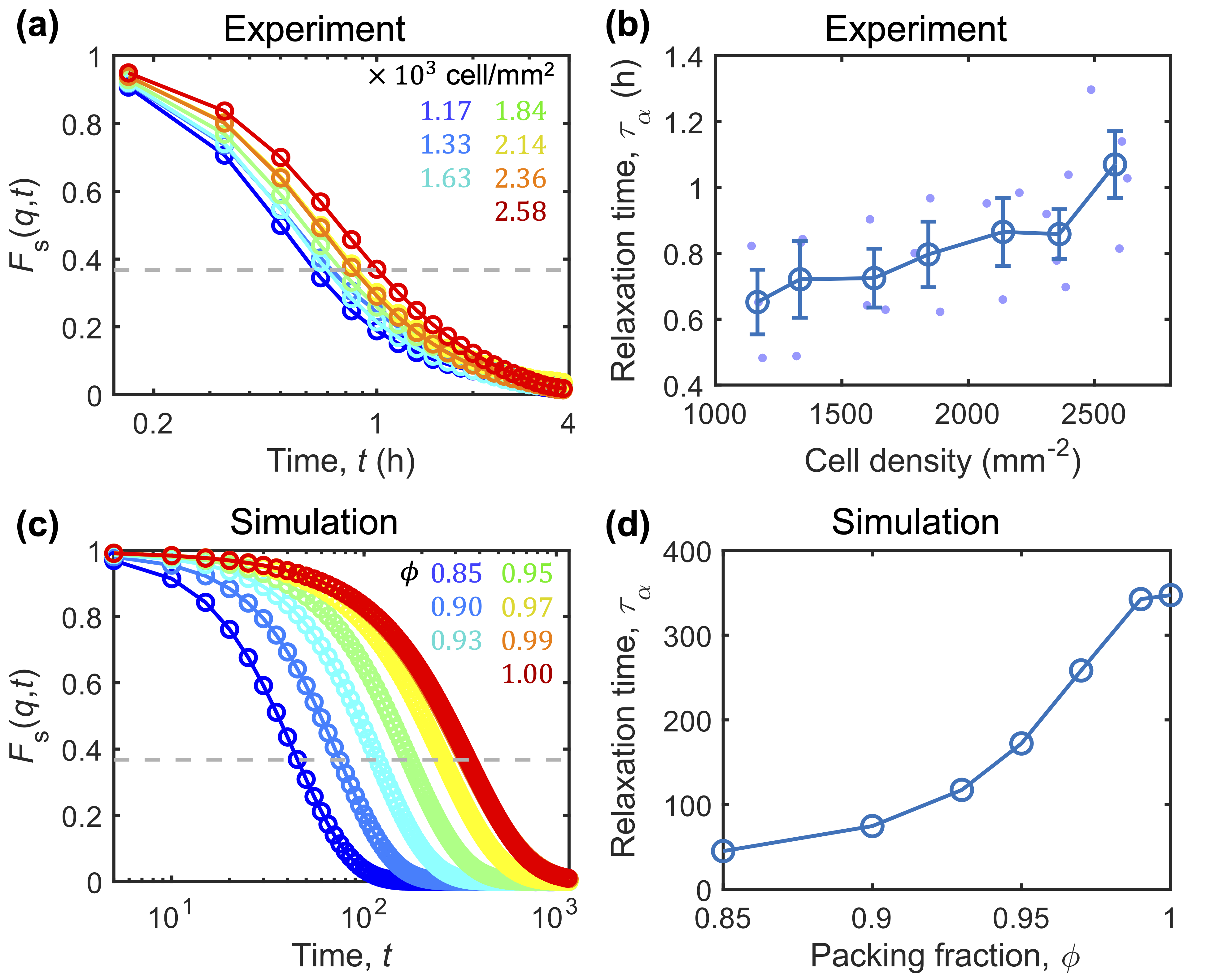}
\caption{
\textbf{Density-dependent structural relaxation in experiments and simulations.}
\textbf{a,} Experimental self-intermediate scattering function, $F_{\mathrm{s}}(q,t)$, at different cell densities, $\rho$. The horizontal dashed line denotes $F_{\mathrm{s}}(q,t)=1/e$.
\textbf{b,} Experimental structural relaxation time, $\tau_\alpha$, defined by $F_{\mathrm{s}}(q,\tau_\alpha)=1/e$, as a function of cell density. Light violet points represent values obtained from individual independent experiments within each density bin, while open circles and error bars indicate the mean and standard error, respectively. 
\textbf{c,} Simulated self-intermediate scattering function at different packing fractions, $\phi$, for $v_0=0.03$ and $\eta=0.01$. The horizontal dashed line denotes $F_{\mathrm{s}}(q,t)=1/e$.
\textbf{d,} Simulated $\tau_{\alpha}$ as a function of packing fraction. Increasing cell density or packing fraction slows structural relaxation in both systems. Nevertheless, all correlation functions shown decay below $1/e$ within the accessible experimental or simulation time window, providing no evidence of complete dynamical arrest under the conditions examined.
\textbf{a, b,} $n\sim2.2\times10^{4}$ and $N_{e}=8$, where $n$ denotes the number of analyzed trajectories, and $N_{e}$ denotes the number of independent experiments.
}
\label{fig:fig3}
\end{figure*}

We characterised collective relaxation using the self-intermediate scattering function $F_{\mathrm{s}}(q,t)$, evaluated at a wave number corresponding to the characteristic intercellular length scale. In the experiments, increasing cell density systematically shifts the decay of the intermediate scattering function $F_{\mathrm{s}}(q,t)$ (see Methods) towards longer times (Fig.~\ref{fig:fig3}\textbf{a}), demonstrating progressively slow cellular rearrangements. 

The structural relaxation time $\tau_\alpha$ was defined by $F_{\mathrm{s}}(q,\tau_\alpha)= 1/e$.
As shown in Fig.~\ref{fig:fig3}\textbf{b}, $\tau_\alpha$ increases significantly with cell density (linear mixed-effects model, $P<$0.001 for the density slope; see Methods). Thus, crowding substantially slows the collective dynamics, but structural relaxation remains observable throughout the investigated density range. Even the densest experimental monolayers therefore retain fluid-like dynamics on our accessible time scale.

The simulations exhibit similar qualitative dependence on crowding. At fixed $v_0=0.03$ and $\eta=0.01$, increasing the packing fraction shifts the decay of $F_{\mathrm{s}}(q,t)$ towards longer times (Fig.~\ref{fig:fig3}\textbf{c}). Correspondingly, $\tau_\alpha$ increases strongly with $\phi$ (Fig.~\ref{fig:fig3}\textbf{d}), with the increase becoming more pronounced above $\phi=0.95$ (approx.) before finally getting saturated at around the highest packing fraction, signifying fluid-like characteristics. 

The experimental cell density and simulated packing fraction cannot be mapped onto one another in a strictly one-to-one manner. Nevertheless, both the experiment and simulation exhibit the same central behaviour: increasing crowding produces a pronounced, progressive slowing of structural relaxation without any evidence of non-relaxing state under the conditions examined. Together with the non-Gaussian displacement statistics shown in Fig.~\ref{fig:fig2}\textbf{e,f}, these results indicate that dense HeLa monolayers display slow, heterogeneous, yet fluid-like collective dynamics that are qualitatively reproduced by the active deformable-cell model.

\begin{figure*}[t]
\centering
\includegraphics[width=0.93\textwidth]{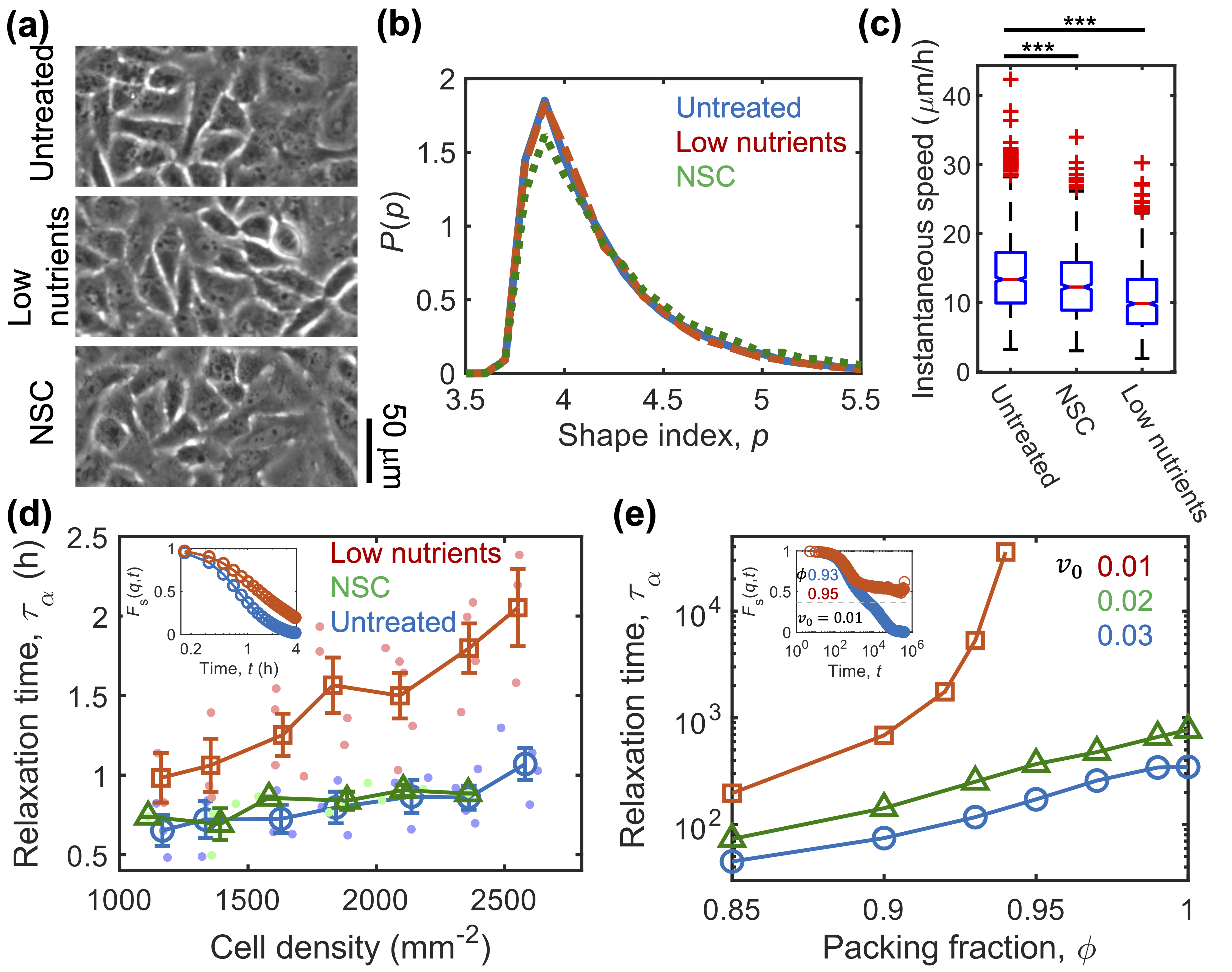}
\caption{
\textbf{Activity-dependent growth of the structural relaxation time.}
\textbf{a,} Phase-contrast images of untreated cells (upper), cells cultured under low-nutrient conditions (middle), and NSC-treated cells (lower). The low-nutrient condition consisted of low-glucose DMEM supplemented with 5$\%$ FBS. Image brightness was adjusted for clarity.
\textbf{b,} Shape-index distributions $P(p)$ of untreated cells (blue), cells cultured under low-nutrient conditions (red), and NSC-treated cells (green).
\textbf{c,} Boxplots of the instantaneous cell speed, $|v(\Delta t)|$, with $\Delta t=10$ min. The Mann–Whitney U test was used to calculate the $P$ values; ***$P<0.001$.
\textbf{d,} Experimental structural relaxation time, $\tau_{\alpha}$, as a function of cell density, $\rho$, for untreated cells (blue), cells cultured under low-nutrient conditions (red), and NSC-treated cells (green). Light points represent values obtained from individual independent experiments within each density bin, while open symbols and error bars indicate the mean and standard error, respectively. Inset: $F_s(q,t)$ for untreated cells and cells cultured under low-nutrient conditions at a cell density of approximately $2.6\times10^{3}$ cells~mm$^{-2}$.
\textbf{e,} Simulation results for $\tau_{\alpha}$ as a function of packing fraction, $\phi$, at self-propulsion speeds $v_0=0.01$ (red squares), $0.02$ (green triangles), and $0.03$ (blue circles), with the shape-deformability parameter fixed at $\eta=0.01$. Open symbols represent the measured state points, and solid lines are guides to the eye. 
Inset: representative $F_s(q,t)$ curves at $\phi=0.93$ and $0.95$ for $v_0=0.01$ and $\eta=0.01$. The horizontal dashed line indicates $F_s(q,t)=1/e$.
\textbf{a--c,} The cell densities were approximately $1.2\times10^{3}$ cells~mm$^{-2}$. \textbf{c,} Untreated, $n=1398$; low nutrients, $n=1002$; NSC-treated, $n=876$. \textbf{d,} Untreated, $n\sim2.2\times10^{4}$, $N_{e}=8$; low nutrients, $n\sim2.4\times10^{4}$, $N_{e}=7$; NSC-treated, $n\sim1.0\times10^{4}$, $N_{e}=8$, where $n$ denotes the number of analyzed trajectories, and $N_{e}$ denotes the number of independent experiments.
}
\label{fig:fig4}
\end{figure*}

Building on the density-dependent slowing established in Fig.~\ref{fig:fig3}, we next asked whether the experimentally observed increase in relaxation time reflects an approach towards dynamical arrest, or whether the HeLa monolayer remains in a liquid-like regime over the experimentally accessible range. To address this question, we compared the relaxation dynamics under different levels of cellular activity in both experiments and simulations (Fig.~\ref{fig:fig4}). 
To relate cellular activity between the two systems, we used the instantaneous cell speed at moderate density as an experimental proxy for cellular activity, since the self-propulsion parameter $v_0$ in the model directly controls the instantaneous speed of individual cells.

We reduced cellular activity experimentally using the Rac1 inhibitor NSC23766 (NSC) or a low-nutrient condition with reduced serum and glucose concentrations. Under both perturbations, cell morphology in the confluent monolayer remained largely unchanged (Fig.~\ref{fig:fig4}\textbf{a}), with all three conditions exhibiting similar shape-index distributions, $P(p)$ (Fig.~\ref{fig:fig4}\textbf{b}). In contrast, NSC treatment moderately but significantly reduced the instantaneous cell speed (Fig.~\ref{fig:fig4}\textbf{c}), consistent with a previous report~\cite{Chen2022Adv.Sci.}. The low-nutrient condition produced a more pronounced decrease in instantaneous speed. Although these perturbations may also affect other cellular properties, including cell deformability, their most prominent effect under the present conditions was a reduction in cell motility.

Under the low-nutrient condition, which exhibited the lowest cell motility, the structural relaxation time $\tau_\alpha$ increased substantially compared with that of untreated cells and showed a stronger dependence on cell density (Fig.~\ref{fig:fig4}\textbf{d}, red symbols). In contrast, Rac1 inhibition had only a modest effect on structural relaxation: despite the reduction in cell speed at moderate density, $\tau_\alpha$ remained close to that of untreated cells at high density (Fig.~\ref{fig:fig4}\textbf{d}, green symbols). In all conditions, $\tau_\alpha$ increased with cell density, with the strongest slowing observed under the low-nutrient condition. Nevertheless, even under this low-motility condition, $F_{\mathrm{s}}(q,t)$ decayed within the observation time and $\tau_\alpha$ remained finite throughout the experimentally accessible density range (Fig.~\ref{fig:fig4}\textbf{d}, inset). Thus, reduced cellular motility can enhance the density-dependent slowing of collective rearrangements, but dense HeLa monolayers remain liquid-like rather than undergoing complete dynamical arrest under the conditions examined.

The simulations provide access to a much broader range of activities and reveal how this liquid-like regime evolves as active motility is further reduced. At a fixed deformability $\eta=0.01$, the structural relaxation time increases with packing fraction for all activities examined (Fig.~\ref{fig:fig4}\textbf{e}). For $v_0=0.03$, corresponding to the representative condition used in Fig.~\ref{fig:fig3}, $\tau_\alpha$ grows continuously with $\phi$ but remains finite even at the highest packing fractions; it saturates at the extreme packing fraction. Reducing the activity to $v_0=0.02$ produces a considerably stronger increase in $\tau_\alpha$, while at the lowest activity, $v_0=0.01$, the growth becomes extremely steep as $\phi$ approaches the densely packed regime.

This distinction is particularly evident from the self-intermediate scattering functions shown in the inset of Fig.~\ref{fig:fig4}\textbf{e}. At $v_0=0.01$ and $\phi=0.93$, $F_{\mathrm{s}}(q,t)$ eventually decays below $1/e$, indicating complete structural relaxation, although on a markedly longer time scale than at larger activities. In contrast, at $\phi=0.95$, the correlation function develops a long-lived plateau and does not cross the $1/e$ threshold within the available simulation time window $[0, 2\times 10^5]$.
The emergence of such persistent relaxation plateaus, together with the rapid growth of $\tau_\alpha$, is consistent with the onset of glass-like dynamical arrest at sufficiently low activity and high packing fraction.

These results identify active motility as a key control parameter governing the dynamical state of the system. Although increasing density results in slow dynamics, under experimentally relevant activity levels, the HeLa monolayer continues to relax and remains liquid-like. In simulations, however, activity can be reduced beyond the experimentally accessed regime, revealing a crossover towards an arrested, glass-like state. This comparison therefore suggests that the absence of glassy arrest in dense HeLa monolayers is not simply a consequence of insufficient crowding; rather, persistent active motility enables structural relaxation and maintains liquid-like collective dynamics even at high cell densities.

\subsection{Comparison of model parameters with experimental estimates}

Finally, we compared the dimensionless parameters used in the simulations with order-of-magnitude estimates based on experimentally reported mechanical and motile properties of adherent cells. 
Since the relevant quantities depend strongly on the cell type and cell state, the following estimates are intended only to provide the correct order of magnitude. 

The Young's modulus of an adherent HeLa cell undergoing slow deformation is typically of the order of
$Y \sim 1~\mathrm{kPa}=10^{3}~\mathrm{N\,m^{-2}}$~\cite{Moeendarbary2013NatureMater}.
A representative value of the effective plasma-membrane tension of an adherent cell is
$\gamma \sim 100~\mathrm{pN\,\mu m^{-1}} =10^{-4}~\mathrm{N\,m^{-1}}$~\cite{Tsujita2021NatCommun}.
As a representative length scale, we take the characteristic cell diameter to be
$L_0=10~\mu\mathrm{m}=10^{-5}~\mathrm{m}$.
Using these quantities, the overlap-energy-density parameter $k$ and the effective cortical-tension parameter $\eta$ are estimated as
\begin{align}
\eta &\sim \gamma L_0,\\
k &\sim YL_0.
\end{align}
The corresponding dimensionless cortical-tension parameter $\tilde{\eta}$ is therefore
\begin{align}
\tilde{\eta} = \frac{\eta}{kL_0} \sim \frac{\gamma}{YL_0} = 0.01,
\end{align}
which lies within the range used in the simulations (see the Method).

To estimate the translational mobility, we use a previously reported cell--substrate friction coefficient $\zeta$, defined as the cell--substrate force per unit cell volume divided by the cell velocity~\cite{McCord2026SoftMatter}.
In that study, the hydrodynamic screening length $\lambda$ and tissue viscosity $\eta_\mathrm{tissue}$ were measured and related to the cell--substrate friction coefficient through $\zeta = \eta_\mathrm{tissue}/\lambda^2$.
Since $\mu_r$ in our model represents the translational mobility of an isolated cell, it can be related to friction coefficient of cell--substrate by
\begin{align}
\mu_r \sim \frac{1}{\zeta V_0} = \frac{\lambda^2}{\eta_\mathrm{tissue}V_0},
\end{align}
where $V_0$ is the cell volume.
For epithelial cells, representative values are
$\eta_\mathrm{tissue}\sim100~\mathrm{Pa\,h}
\sim4\times10^5~\mathrm{Pa\,s}$,
$\lambda\sim200~\mu\mathrm{m}$,
and
$V_0\sim4000~\mu\mathrm{m^3}$~\cite{McCord2026SoftMatter,Zehnder2015BiophysicalJournal}.
These values give
\begin{align}
\mu_r \sim 2.5\times10^1~\mathrm{m\,N^{-1}\,s^{-1}}.
\end{align}
Because these values were obtained for epithelial cells which are different from HeLa cells, we consider only the order of magnitude and adopt $\mu_r \sim 10~\mathrm{m\,N^{-1}\,s^{-1}}$
as a representative value.
The resulting characteristic time scale is
\begin{align}
t_0 = \frac{1}{\mu_r k} \sim \frac{1}{\mu_rYL_0} \sim 10~\mathrm{s}.
\end{align}

Typical migration speed of adherent cells are several to several tens of micrometers per hour.
We adopt a representative HeLa-cell speed of
$v_0=5\times10^{-9}~\mathrm{m\,s^{-1}}$,
corresponding to approximately
$18~\mu\mathrm{m\,h^{-1}}$~\cite{Coene2011JCellBiol}.
The corresponding dimensionless self-propulsion speed is
\begin{align}
\tilde{v}_0 = \frac{v_0t_0}{L_0} = \frac{v_0}{\mu_rYL_0^2} \sim 5\times10^{-3}.
\end{align}
This value is approximately half of that of the lower end of the range used in the simulations, $\tilde{v}_0=0.01$--$0.07$, but remains comparable at the order-of-magnitude level.

Next, we estimate the rotational diffusion coefficient.
The persistence time of adherent-cell migration is typically of the order of several tens of minutes. 
As a representative value, we take $\tau_\mathrm{p}\sim30$--$50~\mathrm{min}\sim10^3~\mathrm{s}$~\cite{Maiuri2015Cell,Ebata2018SciRep}.
Although these values were obtained for cell types different than HeLa cells, they provide a reasonable order-of-magnitude estimate of the persistence time of the self-propulsion direction. 
Assuming that the rotational relaxation time is comparable to the persistence time,
\begin{align}
D_r \sim \frac{1}{\tau_\mathrm{p}} \sim 10^{-3}~\mathrm{s^{-1}}.
\end{align}
The corresponding dimensionless rotational diffusion coefficient is
\begin{align}
\tilde{D}_r = D_rt_0 \sim 0.01,
\end{align}
which agrees with the value used in the simulations.

Finally, we compare the dimensionless structural relaxation times.
In the experiments, the relaxation time ranges from approximately $0.4$ to $1~\mathrm{h}$ (Fig.~\ref{fig:fig3}\textbf{b}).
Using $t_0\sim10~\mathrm{s}$, the corresponding dimensionless relaxation time is therefore,
\begin{align}
\tilde{\tau}_{\alpha} = \frac{\tau_{\alpha}}{t_0} \sim 100\text{--}400,
\end{align}
which is comparable to the typical range of approximately $50$--$400$ obtained in the simulations (Fig.~\ref{fig:fig3}\textbf{d}).

Therefore, although the estimated values depend on several assumptions and representative quantities, all of the estimated dimensionless parameters from experiment remain within approximately one order of magnitude of that of the simulation. This agreement supports the physical plausibility of the parameter ranges adopted in the simulations.

\section{Discussion}

The vertex and self-propelled Voronoi models successfully describe fluid-to-solid transitions with strong cell--cell junctions, where collective dynamics are closely coupled to cell-shape mechanics~\cite{Bi2016Phys.Rev.X,Bi2015NatPhys}. 

Cancer cells, however, often exhibit reduced cell--cell adhesion because of the loss or downregulation of adhesion molecules such as E-cadherin~\cite{Hanahan2000Cell}.  
Therefore, their collective dynamics may differ from those of normal epithelial tissues. Indeed, migration speed in several cancer cell monolayers has been reported to correlate with neither the cell aspect ratio nor the degree of epithelial--mesenchymal transition~\cite{Kim2020BiochemicalandBiophysicalResearchCommunications}. 
Consistent with this observation, HeLa monolayers under low-nutrient conditions exhibited substantially longer relaxation times than those under untreated conditions despite nearly identical cell densities ($\sim2600$ cells/mm$^2$) and median shape indices (3.96 and 3.95, respectively). Although the precise location of the glass transition remains yet to be determined, these results suggest that cell activity, represented by self-propulsion in the present model, is a dominant factor controlling the vitrification of dense HeLa cell collectives.

The correspondence between experimental cell density and simulated packing fraction requires some consideration. In the experiments, neighbouring HeLa cells remain in contact through their lamellipodia even at a density of approximately 1000~cells/mm$^2$. In contrast, the simulations do not explicitly account for the highly flexible and retractable nature of lamellipodia. Experimentally, neighbouring cells can migrate past one another while locally retracting their lamellipodia, suggesting that lamellipodia-mediated contacts provide relatively weak steric hindrance. Strong excluded-volume interactions are therefore expected to emerge only when crowding causes substantial lamellipodial retraction and brings the cell bodies into direct contact.

From images of moderately spread HeLa cells, we estimate the area excluding lamellipodia to be approximately 400--500~$\mu$m$^2$. Although a quantitative mapping remains uncertain, this estimate suggests that the experimental density ($\sim2500$~cells/mm$^2$) corresponds to an effective packing fraction of order unity. The simulated packing fraction should therefore be interpreted as an effective measure based on the sterically interacting cell bodies, rather than the geometric area fraction defined by the total projected cell area.

Adherent cells spontaneously elongate on a substrate even in the absence of external forces, whereas particles in the present model deform only in response to collisional forces. Consequently, isolated cells are circular in the model, and their shape progressively deviates from a circle as the packing fraction increases. In contrast, the median shape index of HeLa cells in the experiments decreases only slightly, from 4.06 to 3.96, as the cell density increases from 1200 to 2600 cells/mm$^2$. This weak density dependence suggests that steric interactions primarily suppress cell elongation rather than inducing deformation. Therefore, to reproduce the experimental observations with more precision, the model should be extended to incorporate active cell deformation driven by intracellular processes rather than solely passive deformation arising from cell--cell collisions~\cite{Ohta2016PhysicaD:NonlinearPhenomena,Ebata2018SciRep}. 

\section{Conclusions} 
In this study, we combined time-lapse imaging of dense HeLa cell monolayers with numerical simulations of a deformable-cell model to examine how crowding and cellular motility regulate collective relaxation. Within the experimentally accessible density and time ranges, increasing cell density substantially slowed collective rearrangements in untreated HeLa monolayers, but structural relaxation remained observable throughout the investigated range. Dense HeLa monolayers therefore exhibited slow, heterogeneous, yet liquid-like collective dynamics under the untreated conditions examined.  
Under low-nutrient conditions, in which the instantaneous cell speed was reduced by approximately 26$\%$ at moderate density, structural relaxation became substantially slower, despite little change in the cell-shape distribution. In contrast, Rac1 inhibition produced a smaller reduction in instantaneous cell speed (approximately 8$\%$) and had only a modest effect on structural relaxation.

The simulations captured the density-dependent slowing observed experimentally and further showed that reducing self-propulsion enhances the growth of the structural relaxation time and promotes long-lived caging dynamics at high packing fractions. Together, these results indicate that persistent cellular motility is an important factor in maintaining structural relaxation in dense HeLa monolayers, while other cellular properties modified under low-nutrient conditions may also contribute to the observed slowing. This behaviour is reminiscent of that of compressible equilibrium glass formers, in which sufficiently high temperatures can prevent glassy arrest even upon compression \cite{Berthier2009EPL, Berthier2009Phys.Rev.E}. In the present active system, self-propulsion similarly acts as a dynamical effective temperature: persistent activity facilitates cage escape and thereby maintains structural relaxation at high density.

The qualitative correspondence between experiments and simulations in cell-shape statistics, displacement distributions, and relaxation dynamics, together with order-of-magnitude estimates of the dimensionless simulation parameters, provides a physical basis for connecting the active deformable-cell model to dense HeLa monolayers. Although a one-to-one mapping between experimental conditions and model parameters remains beyond the scope of the present study, this correspondence provides a useful framework for interpreting the collective dynamics of HeLa cells in terms of physically meaningful model parameters and for identifying mechanisms that control their dynamical slowing.

\section*{Methods}
Here we describe the deformable-cell model and its associated numerical implementation used in the simulations, followed by experimental methods for cell culture, time-lapse imaging, and image analysis.

\subsection*{Deformable-cell model}
We simulated the cancer-cell monolayer using a modified version of the active Fourier-contour deformable-cell model~\cite{Saito2024Sci.Adv.}, in which each cell is represented as an active, continuously deformable, two-dimensional object of different sizes. The boundary of cell $i$, centered at $\mathbf{r}^{\,i}_\mathrm{c}$, is described in polar coordinates by
\begin{equation}
\frac{R_i(\theta)}{R_{0,i}}
=
a_0^{i}
+
\sum_{n=2}^{M}
\left\{
a_n^{i}\cos\left[n(\theta-\theta_i)\right]
+
b_n^{i}\sin\left[n(\theta-\theta_i)\right]
\right\},
\label{eq:fourier_contour}
\end{equation}
where $R_{0,i}$ is the characteristic radius, $\theta_i$ the polarity direction, and $\{a_n^{i},b_n^{i}\}$ time-dependent Fourier coefficients. The first-order coefficients are set to zero, $a_1^i=b_1^i=0$, so that the polar origin is uniquely defined, and the zeroth-order coefficient is fixed by the constant-area constraint, $a_0^i=\sqrt{1-\sum_{n=2}^{M}[(a_n^i)^2+(b_n^i)^2]/2}$. The contour is truncated at $M=6$, giving smooth anisotropic deformations at low computational cost.

Each cell is represented by a smooth scalar field $\varphi_i(\mathbf{r}) = \frac{1}{2}\{1+\tanh([R_i(\hat\theta_i(\mathbf{r}))-\Delta_i(\mathbf{r})]/\epsilon)\}$, with $\Delta_i(\mathbf{r})=|\mathbf{r}-\mathbf{r}^{\,i}_\mathrm{c}|$ and $\hat\theta_i(\mathbf{r})=\arg(\mathbf{r}-\mathbf{r}^{\,i}_\mathrm{c})$, so that $\varphi_i\simeq1$ inside the cell and $\varphi_i\simeq0$ outside, with interface width $\epsilon$. In the sharp-interface limit $\epsilon\to0$, the overlap integral reduces to a one-dimensional contour integral, substantially lowering the numerical cost relative to a conventional multicellular phase-field calculation.

The mechanical energy is $\mathcal{H}=\mathcal{H}_{\mathrm{int}}+\mathcal{H}_{\mathrm{per}}$, where the excluded-volume term
\begin{equation}
\mathcal{H}_{\mathrm{int}}
= k\sum_{i<j}
\int
\varphi_i(\mathbf{r})
\varphi_j(\mathbf{r})
\mathrm{d} \mathbf{r},
\label{eq:overlap_hamiltonian}
\end{equation}
penalizes cell--cell overlap ($k$: overlap-energy density) and produces steric forces, collision-induced torques, and shape-deformation forces, while the perimeter term
\begin{equation}
\mathcal{H}_{\mathrm{per}}=\eta\sum_i l_i,\quad l_i=\int_{-\pi}^{\pi}\sqrt{R_i^{2}(\theta)+\left[
\frac{\partial R_i(\theta)}{\partial\theta}\right]^{2}
}
\mathrm{d}\theta,
\label{eq:perimeter_hamiltonian}
\end{equation}
penalizes contour elongation, with $\eta$ the surface tension parameter and $l_i$ the perimeter of cell $i$; larger $\eta$ suppresses contour elongation, while smaller $\eta$ permits stronger rigidity induced deformation.

The cell centers, polarity angles, and Fourier coefficients evolve via overdamped Langevin dynamics of active particle. Each cell self-propels with velocity $\mathbf{v}_i=v_0(\cos\theta_i,\sin\theta_i)$, where $v_0$ is the self-propulsion speed, giving rise to the equation of motion being,

\begin{equation}
\dot{\mathbf{r}}^{\,i}_\mathrm{c}
=
\mathbf{v}_i
-
\mu_r
\frac{\partial\mathcal{H}}
{\partial\mathbf{r}^{\,i}_\mathrm{c}},
\label{eq:eom_r}
\end{equation}

\begin{equation}
\dot{\theta}_i
=
-
\mu_{\theta}
\frac{\partial\mathcal{H}}
{\partial\theta_i}
+
\sqrt{2D_r}\,\xi_i(t),
\label{eq:eom_theta}
\end{equation}
where $\mu_r$ and $\mu_\theta$ are the translational and rotational mobilities, $D_r$ the rotational diffusion coefficient, $\xi_i(t)$ the Gaussian white noise satisfying $\langle\xi_i(t)\rangle=0$ and $\langle\xi_i(t)\xi_j(t')\rangle=\delta_{ij}\delta(t-t')$. The Fourier coefficients relax down the energy gradient as follows,

\begin{equation}
\dot{a}_n^{i}=-\mu_{ab}\frac{\partial\mathcal{H}}{\partial a_n^{i}},\qquad \dot{b}_n^{i}=-\mu_{ab}\frac{\partial\mathcal{H}}{\partial b_n^{i}},
\label{eq:eom_ab}
\end{equation}

where $\mu_{ab}$ is the shape mobility and $n=2,\ldots,M$; the zeroth-order coefficient $a_0^i$ is not treated as an independent dynamical variable but is instead recomputed after each shape update from the constant-area constraint. Cell collisions can thus simultaneously translate, rotate, and deform cells.

\subsection*{Numerical implementation}

Motivated by the variation of size observed in segmented HeLa-cell microscopy images, we extended the original model by assigning each cell a polydisperse radius
$R_{0,i}= \bar{R}_0(1+\delta_i)$, where $\delta_i\sim\mathcal{N}(0,\sigma_R^2)$, with $\sigma_R=0.10$ and $\mathcal{N}$ the Gaussian distribution. Here only positive radii were accepted. Thus the corresponding preferred area is $A_{0,i}=\pi R_{0,i}^2$ and the packing fraction is $\phi=\sum_{i=1}^{N} A_{0,i}/(L_xL_y)$,
with $L_x$, $L_y$ the simulation-box dimensions.

The model was implemented for simulations in reduced units, with length $L_0=2\bar{R}_0$, energy $E_0=kL_0^2$, and time $t_0=L_0^2/(\mu_rE_0)=1/(\mu_rk)$, where $k$ and $\mu_r$ are defined in Eqs.~\eqref{eq:overlap_hamiltonian} and \eqref{eq:eom_r}; the corresponding reduced parameters in the simulations are $\tilde{v}_0=v_0/(\mu_rkL_0)$ [Eq.~\eqref{eq:eom_r}], $\tilde{\eta}=\eta/(kL_0)$ [Eq.~\eqref{eq:perimeter_hamiltonian}], $\tilde{D}_r=D_r/(\mu_rk)$ [Eq.~\eqref{eq:eom_theta}], $\tilde{\mu}_\theta=\mu_\theta L_0^2/\mu_r$, and $\tilde{\mu}_{ab}=\mu_{ab}L_0^2/\mu_r$ [Eqs.~\eqref{eq:eom_theta} and \eqref{eq:eom_ab}, respectively]. Throughout the text, tildes are omitted, and all physical quantities are expressed in these reduced (dimensionless) units unless otherwise noted. The tildes are reinstated again while explicitly comparing dimensionless simulation parameters with experimental estimates (see Results).

Simulations consist of $N=1024$ cells with radii sampled as mentioned above ($\delta= 10\%$ polydispersity), $M=6$ Fourier modes, and $N_\theta=40$ angular quadrature points; the equations of motion were integrated with time step $\Delta t=0.005$, although tests with $\Delta t=0.001$ also showed numerically converged results. The mobility parameters were fixed at $\mu_\theta=1$, $\mu_{ab}=0.1$, and $D_r=0.01$ (persistence time $D_r^{-1}=100$), while $v_0$, $\eta$, and $\phi$ were scanned over $v_0=0.01$--$0.07$, $\eta=0.01$--$0.08$, and $\phi=0.85$--$1.00$. Periodic boundary conditions were applied in both directions; cell centers were initialized on a nearly hexagonal lattice with polarity angles drawn independently from $\mathrm{Uniform}[0,2\pi)$. Each system was evolved through an initial transient state before configurations and trajectories were recorded for productive statistical analysis.

\subsection*{Cell-shape and dynamical observables}
The instantaneous shape of each cell was quantified using the
dimensionless shape index~\cite{Bi2015NatPhys,Bi2016Phys.Rev.X}
\begin{equation}
p_i
=
\frac{P_i}{\sqrt{A_i}},
\label{eq:shape_index}
\end{equation}
where $P_i=l_i$ is the contour length defined in
Eq.~\eqref{eq:perimeter_hamiltonian} and $A_i$ is the enclosed area of the cell. Distributions of shape index $P(p)$ were computed over all cells and subsequently averaged over the selected steady-state time window.

The self-intermediate scattering function~\cite{Kob1995Phys.Rev.E,Debenedetti2001Nature,Berthier2011Rev.Mod.Phys.,Cavagna2009Phys.Rep.} was calculated as
\begin{equation}
F_s(q,t)
=
\frac{1}{N}
\left\langle
\sum_{j=1}^{N}
\cos
\left[
\mathbf{q}\cdot
\left\{
\mathbf{r}_j(t_0+t)-\mathbf{r}_j(t_0)
\right\}
\right]
\right\rangle_{t_0},
\label{eq:fsqt}
\end{equation}
where the average was taken over time origins $t_0$ and wave vectors
$\mathbf{q}$ of common magnitude $q$. The wave number $q$ was
chosen as the position of the first peak of the structure factor.
The structural relaxation time $\tau_{\alpha}$ was defined through
$F_s(q,\tau_{\alpha})=e^{-1}$.

Single-cell displacement statistics were characterized using the radial
self-part of the van Hove correlation function~\cite{VanHove1954Phys.Rev.,Chaudhuri2007Phys.Rev.Lett.},
\begin{equation}
G_s(r,t)
=
\frac{1}{2\pi r\,N}
\left\langle
\sum_{j=1}^{N}
\delta
\left[
r
-
\left|
\mathbf{r}_j(t_0+t)-\mathbf{r}_j(t_0)
\right|
\right]
\right\rangle_{t_0},
\label{eq:van_hove}
\end{equation}
where $r$ is the magnitude of the single-cell displacement over
lag time $t$, normalized such that
$\int_{0}^{\infty}G_s(r,t)\,\mathrm{d}r=1$.
To remove global translation, trajectories used for
visualization and dynamical analysis were measured relative to the
instantaneous center of mass of the cell population.

\subsection*{Cell culture in experiments}
HeLa cells were cultured in Dulbecco’s modified Eagle’s medium (DMEM; high glucose, 4.5 g/L; Wako) supplemented with 10$\%$ fetal bovine serum (FBS), penicillin (100 U/mL), and streptomycin (0.1 mg/mL) at 37 $^\circ\text{C}$ in a humidified atmosphere containing 5$\%$ CO$_2$.

\subsection*{Time-lapse observation of cell migration in experiments}
Cell migration was monitored using an inverted microscope (TS2R; Nikon, Japan) equipped with a stage-top incubator (WELSX; Tokai Hit, Japan) maintained at 37 $^\circ\text{C}$ in a humidified atmosphere containing 5$\%$ CO$_2$. Prior to time-lapse imaging, cells were seeded onto the fibronectin-coated glass-bottom dishes at densities ranging from 1 $\times$ 10$^4$ to 8 $\times$ 10$^4$ cells/cm$^2$ and incubated overnight in DMEM supplemented with 10$\%$ FBS. Before imaging, the cells were gently washed twice with phosphate-buffered saline (PBS) to remove floating cells and then replenished with fresh DMEM. The cells were subsequently incubated in the stage-top incubator for approximately 3 h to allow stabilization. Phase-contrast images were acquired every 10 min for approximately 24 h using a 10$\times$ (NA = 0.30) Plan Fluor objective lens.

\subsection*{Perturbation of cell migration}
Cell migration at high cell density was reduced by treatment with the Rac1 inhibitor NSC23766 (Focus Biomolecules, United States) or by a low-nutrient condition. For NSC23766 treatment, the culture medium was replaced with DMEM containing 37.5 $\mu$M NSC23766 and 10$\%$ FBS before time-lapse imaging. The cells were then incubated in the stage-top incubator for approximately 3 h to allow the inhibitor to take effect. For analyses of NSC23766-treated cells, only the first 12 h of the time-lapse recordings were used to minimize potential effects of prolonged drug exposure. For the low-nutrient condition, the culture medium was replaced with low-glucose DMEM (1.0 g/L glucose) supplemented with 5$\%$ FBS 1 day before time-lapse imaging, and was replaced again with fresh medium of the same composition approximately 3 h before imaging.

\subsection*{Cell segmentation and tracking in experiments}
Cell migration trajectories and cell shapes were determined using the
deep learning-based segmentation software $\mu$-SAM~\cite{Archit2025NatMethods} together with custom MATLAB routines. To improve segmentation accuracy at high cell
densities, the network weights of $\mu$-SAM were fine-tuned using
manually annotated images. Cell tracking was performed using a custom
MATLAB program: two consecutive segmented images were compared, and
corresponding cells were identified based on the distance between
centroids, image correlation, and the recent trajectory history. Only
trajectories containing at least 24 consecutive time points were retained
for quantitative analysis. The cell trajectories were then smoothed using a five-point moving average to reduce noise, except for the calculation of the van Hove correlation function. 

The shape index, defined by Eq.~\eqref{eq:shape_index}, was calculated from all detected cell contours, which were pooled over the frames used for each analysis.
The self-intermediate scattering function and structural relaxation time were computed from the cell trajectories following Eq.~\eqref{eq:fsqt}. 
For the calculation of $F_s(q,t)$, the wave number was set to $q=2\pi/d$, where $d$ is the mean cell diameter; because $d$ decreased gradually with increasing cell density, the mean value of $d$ over the corresponding 24 time points was used.
We confirmed that $d$ is comparable to the characteristic length scale $2\pi/q_{\rm peak}$, where $q_{\rm peak}$ is the wave number of the first peak of the static structure factor.
The self-part of the van Hove correlation function, $G_s(r,t)$, was calculated according to Eq.~\eqref{eq:van_hove}, with the radial displacement $r$ normalized by the mean cell diameter $d$. 

\subsection*{Structural relaxation analysis in experiments}
To characterize the density dependence of the relaxation time, experiments were performed with several different initial cell densities. Individual-cell trajectories were divided into non-overlapping analysis windows consisting of 24 consecutive time points, corresponding to approximately 4~h of observation. Because cell density gradually increased during the experiments due to cell proliferation, the mean cell density within each analysis window was used to assign trajectories to density bins. 
Cell densities were grouped into bins of width $2.5\times10^2$ cells~mm$^{-2}$, centered at densities ranging from $1.1\times10^3$ to $2.6\times10^3$ cells~mm$^{-2}$.

For each independent experiment, trajectories assigned to the same density bin were used to calculate the self-intermediate scattering function and the corresponding structural relaxation time. The mean and standard error of the relaxation times across independent experiments were then calculated for each density bin.

\subsection*{Statistical analysis}
Differences in instantaneous cell speed between experimental conditions were assessed using the Mann--Whitney $U$ test. To evaluate the overall dependence of the structural relaxation time on cell density while accounting for multiple observations obtained from the same experimental trial, we used a linear mixed-effects model with cell density as a continuous fixed effect. Experimental trial was included as a grouping variable, with random intercepts and random slopes for cell density. A $P$ value $<0.05$ was considered statistically significant. All statistical analyses were performed using MATLAB.

\section{Data and code availability}
The data supporting this study are available within the article and Supporting Information (\hyperref[app:SI]{SI}). Raw data files and numerical/analytical codes supporting this study are publicly available via this link: \url{https://doi.org/10.5281/zenodo.22766980}.

\bibliography{cell_glass}

\section{Acknowledgements}
The authors thank Dr. Kiwamu Yoshii for fruitful discussions. 

\section{Funding}
This work was supported by the JST FOREST Program (JPMJFR212T to T.K.; JPMJFR256N to H.E.), AMED Moonshot Program (JP22zf0127009 to T.K.), JSPS core-to-core program (JPJSCCA20230002 to H.E.), JST ERATO (JPMJER2401 to H.E.), and JSPS KAKENHI (22H04472, 20H05157, 20H00128 to T.K.; 25K07242, 25H01364 to N.S.; 25K00220 to H.E.).

\section{Author contributions}

T.K. and H.E. conceived and supervised the overall project;  S.P., T.K., and H.E. contributed to the conception and designed the research; 
S.P. performed simulation, carried out experiments, and analysed the simulation data; N.S. developed the original deformable cell model and advised S.P. in developing the simulation codes relevant to this research; H.E. designed and carried out experiments and analysed the experimental data; 
S.P., T.K., and H.E. wrote the paper; 
All of the authors interpreted and discussed the results and reviewed the paper.

\section{Competing interests}
The authors declare no competing interests.

\vspace{0.4cm}
\section{Additional information}
\noindent{\bf Supplementary information:} The online version contains supplementary material available at \hyperref[app:SI]{SI}.\\

\noindent{\bf Correspondence} and requests for materials should be addressed to Takeshi Kawasaki and Hiroyuki Ebata.

\clearpage
\onecolumngrid
\setcounter{figure}{0}
\renewcommand{\thefigure}{S\arabic{figure}}
\renewcommand{\theHfigure}{S\arabic{figure}}

\section{Supplementary Information}
\label{app:SI}
\begin{figure*}[ht]
\centering
\includegraphics[width=0.7\textwidth]
{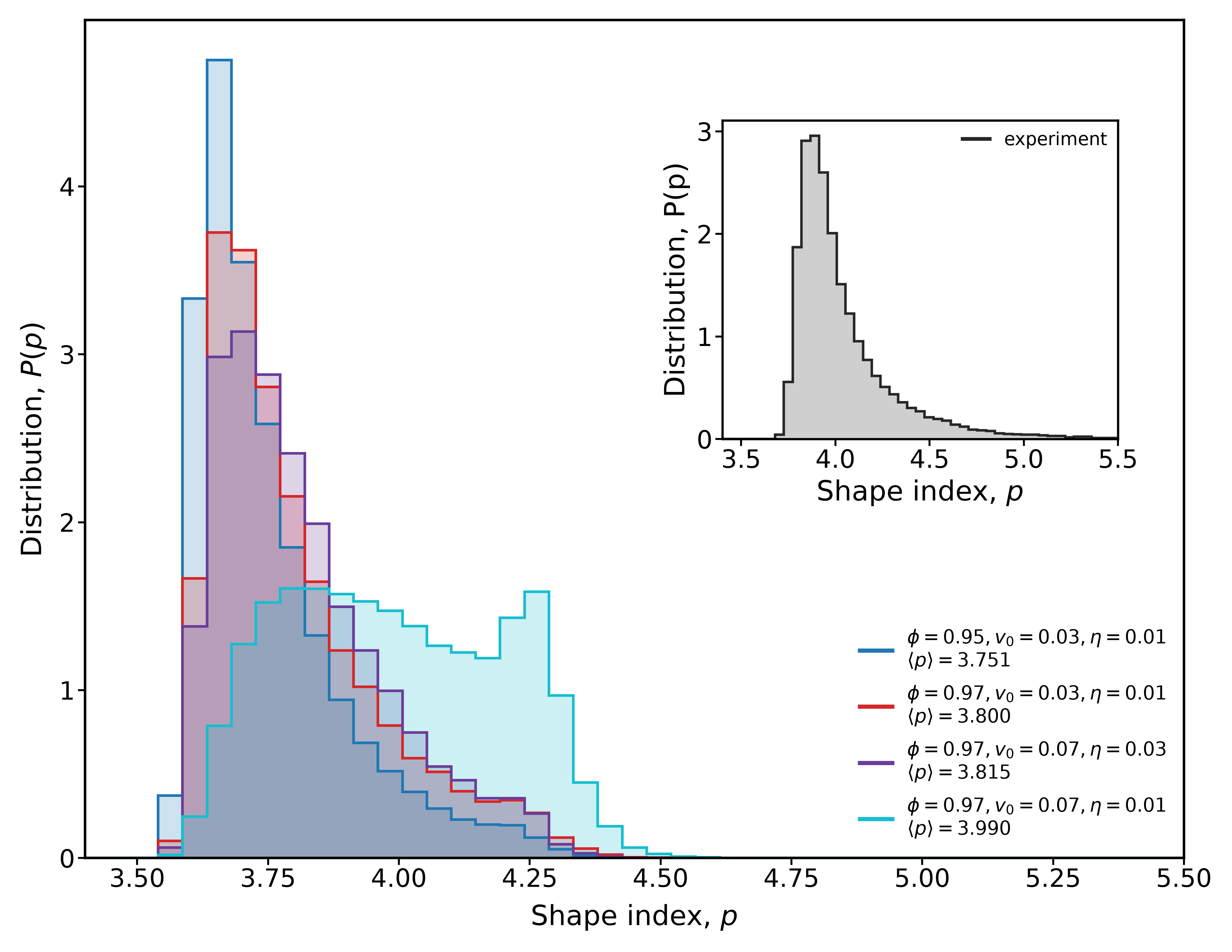}
\caption{
\textbf{Comparison of experimental and simulated shape-index distributions $P(p)$.}
Probability-density distributions of the cell shape index $p$ for four candidate simulation conditions:
$\phi=0.95$, $v_0=0.03$, and $\eta=0.01$
($\langle p\rangle=3.751$);
$\phi=0.97$, $v_0=0.03$, and $\eta=0.01$
($\langle p\rangle=3.800$); and
$\phi=0.97$, $v_0=0.07$, and $\eta=0.03$
($\langle p\rangle=3.815$); and
$\phi=0.97$, $v_0=0.07$, and $\eta=0.01$
($\langle p\rangle=3.990$).
The inset shows the experimental $P(p)$.
The cell density was $2.5\times10^{3}$ cells/mm$^{2}$.
Several simulation conditions produce average shape indices close to the experimental value, although differences remain in the peak position, width, and large-$p$ tail. Considering these distributions together with the displacement statistics in Fig.~\ref{fig:fig2}, the condition $\phi=0.97$, $v_0=0.03$, and $\eta=0.01$ was selected as the representative simulation state for subsequent comparisons.}
\label{fig:supp_shape_distribution}
\end{figure*}

To select a representative simulation condition, we compared the complete
$P(p)$ for four candidate parameter sets with the experimental distribution. As shown in Fig.~\ref{fig:supp_shape_distribution}, the condition $\phi=0.97$, $v_0=0.03$, and $\eta=0.01$ reasonably captures the characteristic range and asymmetric form of the experimental distribution.
Together with the displacement statistics in Fig.~\ref{fig:fig2}, this comparison supports its use as the representative simulation condition.

\end{document}